%% file: main.tex
\documentclass[sigconf, nonacm]{acmart}
\usepackage{stfloats}
\usepackage[a-1b]{pdfx}
\usepackage{float}
\usepackage{enumitem}
\usepackage{tabularx}
\usepackage{listings}
\usepackage[export]{adjustbox}

\usepackage[linesnumbered,ruled,vlined]{algorithm2e}
\usepackage[capitalise]{cleveref}
\usepackage{array}
\newcolumntype{P}[1]{>{\raggedright\arraybackslash}p{#1}} 
\newcolumntype{T}[1]{>{\raggedright\ttfamily\arraybackslash}p{#1}}

\usepackage{graphicx}
\usepackage{subcaption}
\usepackage{microtype}
\usepackage{marginnote} 
\newtheorem{definition}{Definition}

\newcommand{\blue}[1]{\textcolor{black}{#1}}
\renewcommand{\marginnote}[2][]{}
\usepackage{etoolbox}
\BeforeBeginEnvironment{example}{\vspace{-0.5em}}
\AfterEndEnvironment{example}{\vspace{-0.5em}}
\newcommand\vldbdoi{10.14778/3797919.3797940}
\newcommand\vldbpages{1358 - 1371}
\newcommand\vldbvolume{19}
\newcommand\vldbissue{6}
\newcommand\vldbyear{2026}

\newcommand\vldbtitle{\shorttitle} 
\newcommand\vldbavailabilityurl{https://github.com/DBXAI/SysInsight}
\newcommand\vldbpagestyle{empty} 
\newcommand{\sys}{\texttt{SysInsight}\xspace}

\begin{document}

\title{Why Database Manuals Are Not Enough:  Efficient and Reliable Configuration Tuning for DBMSs via Code-Driven LLM Agents}
\settopmatter{authorsperrow=4}
\author{Xinyi Zhang}
\affiliation{%
  \institution{Renmin University, China }
}
\email{xinyizhang.info@ruc.edu.cn}

\author{Tiantian Chen}
\affiliation{%
  \institution{Renmin University, China}
}
\email{chenentt@ruc.edu.cn}

\author{Zhentao Han}

\affiliation{%
    \institution{Renmin University, China}
}
\email{avaleph@ruc.edu.cn}

\author{Zhaoyan Hong}

\affiliation{%
    \institution{Renmin University, China}
}
\email{hongzhaoyan@ruc.edu.cn}

\author{Wei Lu}
\authornote{Corresponding author.}
\affiliation{%
\institution{Renmin University, China}
}
\email{lu-wei@ruc.edu.cn}

\author{Sheng Wang}
\affiliation{%
  \institution{Alibaba Group}
}
\email{sh.wang@alibaba-inc.com}

\author{Mo Sha}
\affiliation{%
  \institution{Alibaba Group}
}
\email{shamo.sm@alibaba-inc.com}

\author{Anni Wang}
\affiliation{%
  \institution{Alibaba Group}
}
\email{wanganni.wan@alibaba-inc.com}

\author{Shuang Liu}

\affiliation{%
    \institution{Renmin University, China}
}
\email{shuang.liu@ruc.edu.cn}

\author{Yakun Zhang}

\affiliation{%
    \institution{Harbin Institute of Technology, Shenzen}
}
\email{zhangyk@hit.edu.cn}




\author{Feifei Li}
\affiliation{%
  \institution{Alibaba Group}
}
\email{lifeifei@alibaba-inc.com}

\author{Xiaoyong Du}
\affiliation{%
\institution{Renmin University, China}
}
\email{duyong@ruc.edu.cn}

\begin{abstract}
\input{0abstract}
\end{abstract}

\maketitle

\pagestyle{\vldbpagestyle}
\begingroup\small\noindent\raggedright\textbf{PVLDB Reference Format:}\\
Xinyi Zhang, Tiantian Chen, Zhentao Han, Zhaoyan Hong, Wei Lu, Sheng
Wang, Mo Sha, Anni Wang, Shuang Liu, Yakun Zhang, Feifei Li, Xiaoyong Du. \vldbtitle. PVLDB, \vldbvolume(\vldbissue): \vldbpages, \vldbyear.\\
\href{https://doi.org/\vldbdoi}{doi:\vldbdoi}
\endgroup
\begingroup
\renewcommand\thefootnote{}\footnote{\noindent
This work is licensed under the Creative Commons BY-NC-ND 4.0 International License. Visit \url{https://creativecommons.org/licenses/by-nc-nd/4.0/} to view a copy of this license. For any use beyond those covered by this license, obtain permission by emailing \href{mailto:info@vldb.org}{info@vldb.org}. Copyright is held by the owner/author(s). Publication rights licensed to the VLDB Endowment. 
\\
\raggedright Proceedings of the VLDB Endowment, Vol. \vldbvolume, No. \vldbissue\ %
ISSN 2150-8097. \\
\href{https://doi.org/\vldbdoi}{doi:\vldbdoi} \\

}\addtocounter{footnote}{-1}\endgroup

\ifdefempty{\vldbavailabilityurl}{}{
\vspace{.3cm}
\begingroup\small\noindent\raggedright\textbf{PVLDB Artifact Availability:}\\
The source code, data, and/or other artifacts have been made available at \url{\vldbavailabilityurl}.
\endgroup
}

\input{1intro}

\input{2pre}

\input{3overview}
\input{4extraction}

\input{5tuning}
\input{6exp}

\input{7conclusion}
\input{8ack}

\bibliographystyle{ACM-Reference-Format}
\balance
\bibliography{sample}
\balance
\end{document}

%% file: 0abstract.tex
Modern database management systems (DBMSs) expose hundreds of configuration knobs that critically influence performance. Existing automated tuning methods either adopt a data-driven paradigm, which incurs substantial  overhead, or rely on manual-driven heuristics extracted from database documentation,  which are often limited 
and overly generic. 
Motivated by the fact that the control logic of configuration knobs is inherently encoded in the DBMS source code, we argue that promising tuning strategies can be mined directly from the code, uncovering fine-grained insights grounded in system internals.
To this end, we propose \sys, a code-driven database tuning system that automatically extracts fine-grained tuning knowledge from DBMS source code to accelerate and stabilize the tuning process. \sys combines static code analysis with LLM-based reasoning to identify knob-controlled execution paths and extract semantic tuning insights. These insights are then transformed into  quantitative and verifiable tuning rules via association rule mining grounded in tuning observations. During online tuning, system diagnosis is applied to identify critical knobs, which are  adjusted under the rule guidance. Evaluations demonstrate that  compared to the  SOTA baseline, \sys converges to the best  configuration on average 7.11$\times$ faster while achieving a 19.9\% performance improvement.



%% file: 1intro.tex
\section{Introduction}

Modern database management systems (DBMSs) expose hundreds of configuration knobs~\cite{duan2009tuning}, and setting suitable values for these knobs is crucial for achieving high throughput and low latency.
However, determining the optimal configurations is an NP-hard problem~\cite{sullivan2004using}.
Traditionally, database administrators (DBAs) invest significant effort in manually tuning these knobs for specific workloads.
To alleviate this burden, many recent studies have focused on automating knob tuning using machine learning (ML) techniques.

Depending on the source of knowledge they utilize, we categorize existing ML-based tuning approaches into two broad paradigms:
(1) \textit{data-driven} methods, which interact iteratively with the DBMS to collect performance data under various configurations and adapt tuning decisions based on the feedback; and
(2) \textit{manual-driven} methods, which extract tuning guidance from database manuals, online technical blogs or Q\&A forums.
\textit{Data-driven} methods~\cite{DBLP:conf/vldb/AgrawalCKMNS04,DBLP:conf/vldb/WeikumMHZ02,DBLP:conf/vldb/StormGLDS06,DBLP:conf/vldb/ShashaB02,DBLP:conf/vldb/ChaudhuriW06,DBLP:conf/vldb/ShashaR92,DBLP:journals/dpd/KossmannS20,DBLP:conf/sigmod/MaAHMPG18, DBLP:journals/corr/abs-2203-14473}  formulate tuning as a  black-box optimization problem, which explores the search space via trial-and-error.
In the absence of effective guidance to prune or prioritize configurations, these methods must iteratively probe the  search space, leading to substantial overhead due to extensive workload replays to gather performance observations.

\textit{Manual-driven} methods~\cite{DBLP:conf/sigmod/Trummer22,DBLP:journals/pvldb/LaoWLWZCCTW24} have recently been proposed to leverage pretrained language models to extract expert heuristics from textual documentation for database tuning.
For example, GPTuner~\cite{DBLP:journals/pvldb/LaoWLWZCCTW24} uses large language models (LLMs)  to extract tuning ranges and recommended values. 
However, the coverage of such heuristics is limited, as not all knobs have expert recommendations recorded in manuals.
In GPTuner, for instance, fewer than 50\% of the knobs are identified to have meaningful ranges smaller than their default ranges (Section~\ref{sec:exp-eff}).
For newly developed or poorly documented systems, the tuning heuristics may be entirely absent, rendering \textit{manual-driven} approaches ineffective.
Second, the information extracted by existing methods is often oversimplified, which lacks fine-grained guidance.
For example,  \texttt{innodb\_buffer\_pool\_size} is suggested to be set to 70\%–80\% of the machine’s physical memory ~\cite{DBLP:conf/sigmod/Trummer22,DBLP:journals/pvldb/LaoWLWZCCTW24}.
However, setting the optimal value for a particular workload still relies on trial-and-error within this range, and may even fall outside it, since these ranges are summarized for common scenarios rather than tailored to specific workloads (Section~\ref{sec:exp-case}).
\begin{figure*}[t]
    \centering
    \scalebox{0.8}{
    \includegraphics{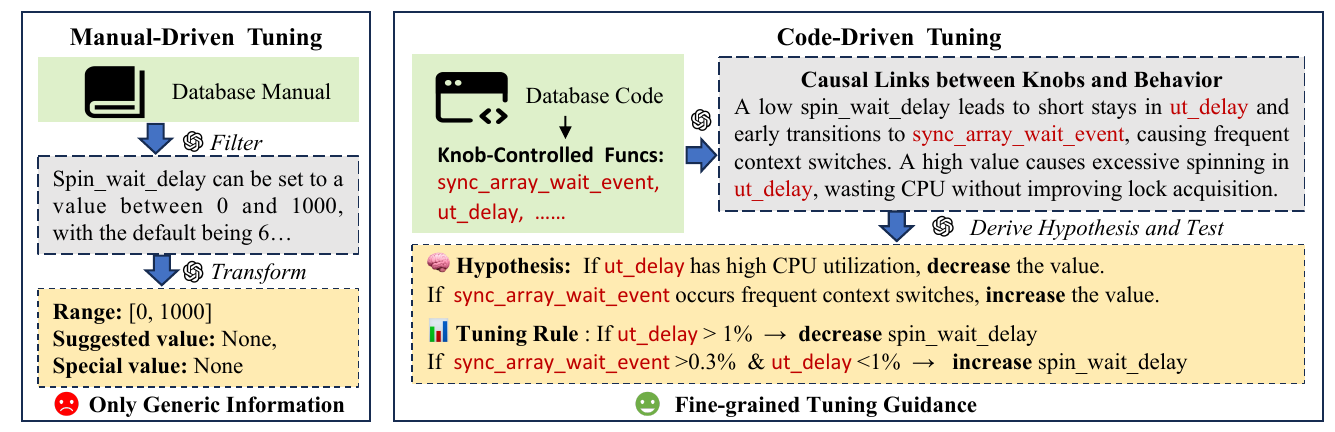}}\vspace{-0.5em}
    \caption{Motivating Example.  
   Manual-driven approach only provides generic information for knob \texttt{spin\_wait\_delay} (i.e., \texttt{innodb\_spin\_wait\_delay}), requiring further trial-and-error.
    \sys addresses this gap by analyzing the knob-controlled functions  (e.g., ut\_delay, sync\_array\_wait\_event) to derive fine-grained tuning guidance rooted in system internals.}
    \label{intro}
    \vspace{-0.5em}
\end{figure*}

\textbf{Motivation.} 
Beyond relying on existing manuals, in this paper, we argue that  tuning strategies can be derived from the dedicated system code itself, as configuration knobs fundamentally  influence performance  by controlling predefined execution paths (Section~\ref{sec:pre-knob}).
From this perspective, we propose a new \textit{code-driven} tuning paradigm, which we instantiate in a system called \sys, designed to act as a \underline{Sys}tem architect with deep \underline{Insight} into the database's internal mechanisms. 
\sys enables precise configuration tuning by identifying runtime performance bottlenecks, analyzing the relevant system code to locate the controlling knobs,  and tuning them based on their impact on execution paths.

To illustrate how code-driven tuning differs from existing manual-driven methods, Figure~\ref{intro} compares \sys\ with GPTuner~\cite{DBLP:journals/pvldb/LaoWLWZCCTW24}, using \texttt{innodb\_spin\_wait\_delay} as an example.
GPTuner fails to obtain compact ranges (the 0–1000 range merely reflects MySQL’s default bounds) or reliable recommended values due to the lack of expert heuristics and the absence of widely accepted tuning guidelines for \texttt{innodb\_spin\_wait\_delay} available in the manual.
In contrast, \sys\ analyzes the database source code to identify the functions affected by \texttt{innodb\_spin\_wait\_delay}, finding its control over the trade-off between \texttt{ut\_delay} and \texttt{sync\_array\_wait\allowbreak\_event}.
Based on these causal links, \sys formulates  hypotheses and refines them into actionable tuning rules through empirical validation, uncovering fine-grained tuning guidance beyond what is provided in database manuals.

There are several challenges in achieving automatic \textit{code-driven} tuning.
First, while the effects of knobs are encoded in the DBMS codebase, analyzing their control logic and associated behavioral patterns is  highly complex and labor-intensive.
Recent advances in  LLMs have demonstrated strong capabilities in code generation and reasoning~\cite{DBLP:journals/corr/abs-2406-00515,DBLP:conf/icse/NamMHVM24}.
However, modern DBMS codebases often exceed millions of lines, making it difficult to isolate how a knob influences execution (e.g., which functions or loops it controls) without overwhelming the LLM or losing context. 
Moreover, existing code summarization techniques~\cite{DBLP:conf/icse/SunMLZFLDLC25} are primarily designed for functional documentation rather than identifying knob-specific control paths.
Thus, a challenge lies in automatically distilling tuning-relevant insights such as high-level control dependencies and performance implications  from the large and complex codebase (\textbf{C1}).

Second, even when such insights are available, converting them into \textit{reliable tuning guidance} remains difficult (\textbf{C2}). 
With hundreds of knobs, it is often unclear which ones are more likely to yield performance improvements.
Even for a promising knob, determining the adjustment direction 
(increase or decrease) and step size is non-trivial. 
For example, although we may know that a high \texttt{ut\_delay} suggests 
decreasing \texttt{innodb\_spin\_wait\_delay}, this insight is insufficient 
without defining a safe range and precise conditions. 
Effective guidance must specify both the triggering conditions and the  adjustment to avoid aggressive changes that might downgrade system  performance; for instance, ``if the \texttt{ut\_delay} ratio exceeds a certain  threshold, decrease \texttt{innodb\_spin\_wait\_delay} by a certain  step until the condition is no longer met''.
Crafting such guidance is difficult as it requires precisely specifying conditions, directions, and steps under diverse contexts without introducing instability.

Third, the absence of a robustness verification mechanism presents a critical challenge, particularly for mitigating issues such as hallucinations in LLMs~\cite{DBLP:journals/tois/HuangYMZFWCPFQL25}.
Unlike general-purpose text generation, knob tuning operates in high-stakes environments where each recommendation can directly impact throughput, latency, or resource contention.
However, existing approaches---whether data-driven or manual-driven—--often explore a wide range of suboptimal or even harmful configurations during tuning, as they lack safeguards to assess the trustworthiness of the tuning logic beforehand (Section~\ref{sec:exp-reliable}).
To ensure safety and effectiveness, tuning decisions must be grounded in real system behavior and verifiable prior to deployment.
Yet both data-driven models and LLMs offer no intrinsic guarantees of correctness or generalizability, and tuning outcomes are highly context-dependent.
As a result, assessing whether a tuning knowledge is  reliable remains an open and difficult problem, yet essential for trustworthy tuning outcomes (\textbf{C3}).

\sys  addresses these challenges with the following  designs.
First, to uncover knob effect from large-scale codebase, we propose a code-driven tuning knowledge extraction approach that combines static analysis and LLM-based code reasoning.
Static analysis localizes functions affected by target knobs to reduce irrelevant code exploration, while an LLM agent explores their surrounding context (e.g., callees and class definitions) to formulate potential tuning hypotheses based on their impact on system behavior (for \textbf{C1}).
Second, to transform these semantic tuning hypotheses into effective and reliable tuning guidance, we introduce a rule induction framework that uncovers knob adjustments consistently associated with performance improvements under similar runtime contexts (for \textbf{C2}).
Third, we propose a reliability verification mechanism that maintains rule-level confidence scores based on performance feedback, which  supports continuous rule refinement and safe reuse (for \textbf{C3}).
Finally, we integrate these components into a code-driven, rule-augmented tuning framework which identifies bottleneck functions, retrieves relevant rules, and suggests adjustments for the associated knobs. 
The main contributions  are summarized as follows:
\begin{itemize}[leftmargin=*]
\item 
We leverage database source code to trace knob-to-function influence paths, identify critical knobs, and uncover promising tuning strategies.
To the best of our knowledge, this work presents the first automatic code-driven configuration tuning paradigm.

\item  
We define context-aware and verifiable tuning rules and develop a customized association rule mining pipeline that transforms promising tuning strategies into reliable, quantitative adjustments grounded in empirical observations.
\item 
We integrate code-driven  extraction and rule induction into a unified framework that adaptively retrieves, verifies, and refines tuning rules, enabling efficient and reliable knob tuning.

\item  We  conduct extensive experiments  to validate the effectiveness  of \sys, which on average converges to the best configuration 7.11$\times$ faster, while achieving a 19.9\% performance improvement compared to the second-best baseline.
\end{itemize}

%% file: 2pre.tex
\section{Preliminary}

\subsection{Database Knob Tuning}
Consider a database system with a set of configuration knobs denoted by \(\{\theta_1, \dots, \theta_m\}\). We define its configuration space as
\(
\boldsymbol{\Theta} = \Theta_1 \times \cdots \times \Theta_m,
\)
where each \(\Theta_i\) denotes the domain of knob \(\theta_i\),  which can be either continuous or categorical.
We denote the database context metrics as \(\mathcal{C} = \{\mathcal{R}, \mathcal{W}\}\), where \(\mathcal{R}\) represents runtime metrics (e.g., function sampling rates, internal counters), and \(\mathcal{W}\) denotes workload characteristics. Let
\(
f
\)
be the performance metric to optimize. Given a specific configuration \(\boldsymbol{\theta}\in \boldsymbol{\Theta}\) and context metrics \(c\in\mathcal{C}\), the corresponding performance \(f(c, \boldsymbol{\theta})\) can be observed after evaluation in the database system.
Assuming the objective is a maximization problem, the goal of database knob tuning is to identify an optimal configuration \(\boldsymbol{\theta}^*\in\boldsymbol{\Theta}\) that maximizes performance under context \(c\), i.e.,
\begin{equation}
\boldsymbol{\theta}^{*} = \underset{\boldsymbol{\theta}\in \boldsymbol{\Theta}}{\arg\max}\, f(c, \boldsymbol{\theta}).
\end{equation}

\noindent To tackle this problem, existing efforts mainly focus on two dimensions: knob selection  and configuration recommendation:

\textbf{Knob Selection.}\label{sec:pre1}
Modern database systems expose hundreds of knobs, but their performance impact can vary significantly across different workloads. The goal of knob selection is to identify which knobs need to be tuned for the current workload. We discuss existing approaches from the two perspectives:
\begin{itemize}[leftmargin=*]
\item \textit{Data-driven} methods~\cite{DBLP:conf/sigmod/AkenPGZ17, DBLP:conf/sigmod/CaiLZZZLLCYX22,DBLP:journals/corr/abs-2001-08002, 10.14778/3632093.3632114}  sample a large number of observations under different configurations  and fit a ML model based on the observations to  rank the knobs in terms of their importance measurement~\cite{tibshirani1996regression, DBLP:journals/bioinformatics/NembriniKW18, DBLP:journals/corr/abs-2011-15001}.
However, prior studies~\cite{DBLP:conf/hotstorage/KanellisAV20} have shown that accurately identifying important knobs requires substantial sampling cost, making it inefficient in practice.
\item \textit{Manual-driven} methods~\cite{DBLP:journals/pvldb/LaoWLWZCCTW24, DBLP:journals/corr/abs-2408-02213} rely  on LLMs to suggest important knobs.
These methods prompt the LLM with workload descriptions (e.g., OLTP vs. OLAP) and query plans, assuming that its pre-training on manuals and forum discussions enables effective recommendations.
However, such methods tend to favor popular knobs frequently seen in the pre-trained data of LLMs while overlooking critical but under-documented ones.
\end{itemize}

\textbf{Configuration Recommendation.}
Once the tuning knobs are selected, the next step is to recommend their configuration values.

\begin{itemize}[leftmargin=*]
\item  \textit{Data-driven} methods works iteratively:  applies a suggested configuration to the DBMS, and  update their internal model or strategy based on the observed performance. 
Depending on the underlying strategy, these methods can be further categorized into:
(1) Heuristic-based methods~\cite{MySQLTuner, pgtune, zhu2017bestconfig,DBLP:conf/IEEEpact/AnselKVRBOA14}, which rely on rule-based or search-based heuristics to explore configuration space;
(2) Bayesian Optimization (BO)-based methods~\cite{DBLP:journals/pvldb/DuanTB09, DBLP:conf/sigmod/AkenPGZ17, DBLP:journals/pvldb/CeredaVCD21, DBLP:journals/pvldb/KanellisDKMCV22,DBLP:journals/corr/abs-2203-14473,DBLP:journals/pvldb/CeredaVCD21,10.14778/3632093.3632114}, which model the configuration–performance relationship using a probabilistic surrogate and select the next configuration by maximizing an acquisition function;
(3) Reinforcement Learning (RL)-based methods~\cite{DBLP:conf/sigmod/ZhangLZLXCXWCLR19, DBLP:conf/sigmod/CaiLZZZLLCYX22,DBLP:journals/pvldb/LiZLG19,DBLP:journals/jcst/GeCC21,DBLP:journals/pvldb/WangTB21}, which treat tuning as a sequential decision-making process and learn  policies through interactions with  DBMSs.

\item \textit{Manual-driven} methods~\cite{DBLP:conf/sigmod/Trummer22,DBLP:journals/pvldb/LaoWLWZCCTW24} further accelerate configuration search by extracting expert knowledge from manuals or online content to prune the search space.
For example, DB-BERT~\cite{DBLP:conf/sigmod/Trummer22} transforms textual descriptions into tuning hints of the form $P = f(v, S)$, where $v$ is a candidate configuration, $S$ encodes hardware properties (e.g., memory size) and $P$ is the suggested value given $v$ and $S$ (e.g., 70\% of the instance memory). 
However, these recommendations typically consider only  hardware factors and do not adapt to workload characteristics, requiring further RL-based exploration to refine configurations.
GPTuner~\cite{DBLP:journals/pvldb/LaoWLWZCCTW24} extracts  min/max bounds, default, suggested, and special values for each knob   and then applies BO to  explore a reduced and discretized search space defined by these ranges and values.

\end{itemize}




\begin{figure*}[!t]
  \centering
  \begin{subfigure}[t]{0.48\linewidth}
    \centering
    \includegraphics[width=\linewidth]{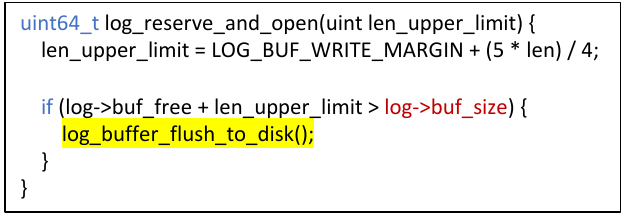}
    \caption{Explicit Control-Dependent on \texttt{innodb\_log\_buffer\_size}.}
    \label{fig:control1}
  \end{subfigure}
  \hfill
  \begin{subfigure}[t]{0.48\linewidth}
    \centering
    \includegraphics[width=\linewidth]{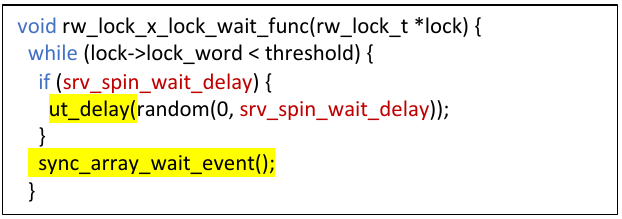}
    \caption{Implicit Control-Dependent on \texttt{innodb\_spin\_wait\_delay}.}
    \label{fig:knob2}
  \end{subfigure}
\vspace{-0.5em}
  \caption{\marginnote{\blue{R2O4}}Simplified Code Snippets Showing How Knobs Influence System Behavior: 
 \blue{(a) \texttt{innodb\_log\_buffer\_size}, implemented internally via \texttt{log->buf\_size}, determines whether a log buffer flush is triggered. 
(b) \texttt{innodb\_spin\_wait\_delay}, implemented internally via \texttt{srv\_spin\_wait\_delay}, controls the polling frequency in the spin-lock loop.}}
  \label{fig:code-snippets}
  \vspace{-0.5em}
\end{figure*}

\subsection{ Knob Impacts on DBMS Behaviors}\label{sec:pre-knob}

We analyze how knobs influence
DBMS behaviors by controlling the execution paths defined in the system code, and further analyze how this influence can be leveraged for performance tuning.

\textbf{Explicit Control}: Configuration knobs are commonly used as switches in software, enabling users to adjust various functional features of the system. 
In DBMS, knobs often appear in conditional statements (e.g., if, switch, for, and while), and determine which branch of the program will be executed. As a result, the dominated branches and blocks are control-dependent on the  knob.

\begin{example}
\texttt{innodb\_log\_buffer\_size} determines the size of the buffer that InnoDB uses to write to the log files on disk.
 As shown in Figure \ref{fig:control1}, if the \texttt{buf\_size} is smaller than the free size plus the length of new log, MySQL will trigger a costly synchronous buffer flush operation. This is an explicit control. 
 If there has transactions with large blob/text fields, the buffer can fill up very quickly and incur performance hit.
\end{example}

\textbf{Implicit Control.} 
Beyond explicit control flow defined by conditional branches, configuration knobs can also implicitly propagate control dependency. 
For instance, if a knob influences the parameter of a delay operation (e.g., a \texttt{sleep} call) within a loop, then all statements inside the loop are implicitly control-dependent on that knob.
This is because  the knob  alters the execution frequency and timing, thereby indirectly affecting control behavior.

\begin{example}
 As shown in Figure~\ref{fig:knob2}, when 
\texttt{srv\_spin\_wait\_delay} is non-zero, \texttt{ut\_delay()} injects a random  delay (from 0 to \texttt{srv\_spin\allowbreak\_wait\_delay}) before rechecking the lock condition.
A larger \texttt{innodb\allowbreak\_spin\_wait\_delay} value increases the delay in each iteration, reducing the frequency of calls to \texttt{sync\_array\_wait\_event()}. 
Therefore, all basic blocks 
inside the \texttt{while} loop are implicitly control-dependent on 
\texttt{innodb\_spin\_wait\_delay}.
\end{example}

\textbf{Tuning Hypothesis.}
After analyzing the impacts of a knob on DBMS behaviors, one can formulate its potential tuning strategies, such as adjustment directions given inferred from the observed state of the associated functions.   
These strategies are promising but not yet verified. Thus, we refer to them as tuning hypotheses.
A tuning hypothesis semantically provides an initial adjustment direction based on profiling signals of the functions a knob controls.

\begin{example}
Given the control effect of \texttt{innodb\_log\_buffer\allowbreak\_size}, one can hypothesize that frequent execution of \texttt{log\_buffer\allowbreak\_flush\allowbreak\_to\_disk()} suggests increasing \texttt{innodb\_log\_buffer\_size}.
Similarly, given the control effect of \texttt{innodb\_spin\_wait\_delay}, one can hypothesize that:
If \texttt{ut\_delay()} consumes excessive CPU time, lowering \texttt{innodb\_spin\_wait\_delay} is recommended to reduce spin overhead; conversely, if \texttt{sync\_array\_wait\_event()} dominates, increasing the delay helps avoid frequent context switches. 
\end{example}

\blue{\marginnote{\textbf{R4W4}}Although the tuning hypotheses are formulated on a per-knob  basis, the presence of co-triggered bottleneck functions can implicitly reflect inter-knob dependencies. 
For example, frequent invocation of \texttt{buf\_LRU\_get\_free\_block} and \texttt{buf\_flush\_sync\_all\_buf\_pools} 
indicate pressure on both memory and log subsystems, suggesting that both 
\texttt{innodb\_buffer\_pool\_size} and \texttt{innodb\_log\_file\_size} should be enlarged.}


%% file: 3overview.tex
\begin{figure*}[t]
    \centering
    \scalebox{0.7}{
    \includegraphics{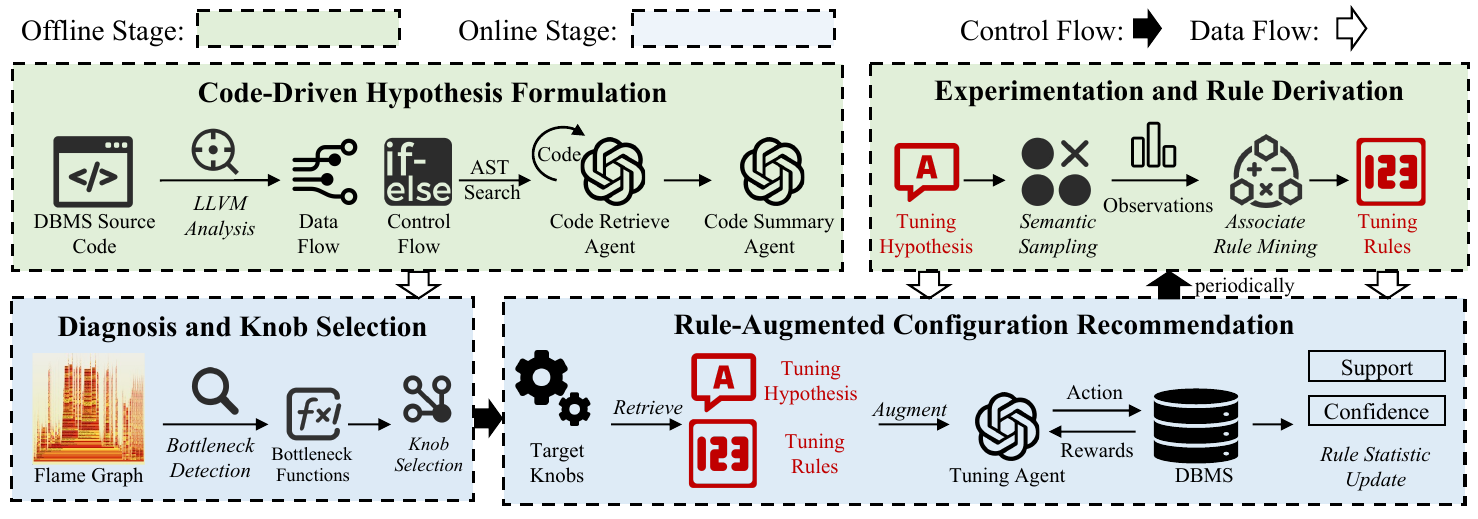}}
    \caption{\marginnote{\blue{R4O3}}\blue{\sys Workflow.}}
    \label{fig:overview}
    \vspace{-0.3em}
\end{figure*}
\section{System Overview}\label{sec:overview}
\sys  emulates the action of a DBMS expert who identifies system issues, examines related code, performs targeted tuning, and summarizes tuning experience for future reuse. 
\blue{\marginnote{\textbf{R4O3}}Figure~\ref{figR:overview} illustrates the workflow of \sys, which consists of both online and offline phases. 
The offline phase builds a reliable repository of code-driven tuning knowledge (Section~\ref{sec:handel}). 
During online tuning, bottleneck functions are identified and their associated knobs are tuned under the guidance of the learned knowledge (Section~\ref{sec:frame}).}

\noindent\textbf{Code-Driven Hypothesis Formulation.} 
This process leverages the extracted control and data flow  to trace how the knob’s value propagates through the program logic and impacts system execution.
A code retrieval agent gathers relevant unresolved context via Abstract Syntax Tree (AST) search.
Based on the context, a summary agent formulates semantic tuning hypotheses expected to improve system performance.

\noindent\textbf{Experimentation and Rule Derivation.}
Based on the semantic tuning hypotheses, \sys conducts tuning experiments by sampling configurations that conform to these hypotheses.
To further distill precise and actionable tuning guidance from the experimental observations, association rule mining techniques are applied to extract tuning rules that explicitly specify the conditions under which knob adjustments should be triggered, as well as the direction and step size of each adjustment.

\noindent\textbf{System Diagnosis and Knob Selection.} 
During tuning, \sys collects runtime metrics and detects the deviations from expected performance to locate bottleneck functions.
To identify relevant configuration knobs, \sys then performs static analysis on the DBMS source code, tracing the control and data flows to construct function–knob mappings, allowing \sys to pinpoint the knobs that directly or indirectly affect the identified bottlenecks.




\noindent\textbf{Rule-Augmented Configuration Recommendation.}
 \sys leverages the derived tuning rules in conjunction with semantic hypotheses to guide online configuration tuning. 
It retrieves  relevant rules whose antecedent predicates match the current runtime metrics, ranking them by expected improvement to prioritize the most promising candidates and accordingly suggests a configuration.
After applying the suggested configuration, \sys observes the resulting performance  and  updates rule statistics to track the reliability of each rule.
\blue{As tuning observations accumulate, \sys periodically applies
the association rule mining to induce  tuning rules.}

%% file: 4extraction.tex
\section{Code-Driven Knowledge  Handler}\label{sec:handel}

\subsection{Function-Knob Mapping Construction}\label{sec:sem1}
To accurately determine how configuration knobs influence system execution while avoiding excessive input to the LLM, \sys first constructs a function–knob mapping that identifies the functions controlled by each knob. 
This mapping is essential for linking configuration changes to concrete execution paths and for guiding subsequent hypothesis formulation and knob selection.

Given a target knob $k$, \sys first leverages ConfMapper~\cite{DBLP:conf/qrs/ZhouLLDLX16} to locate the initial program variable $v^k$ to which the knob configuration is mapped to. 
Then,  it adopts static taint analysis  technique~\cite{wang2023conftainter} to track both data-flow and control-flow propagation from $v^k$ by analyzing the LLVM Intermediate Representation (IR)~\cite{DBLP:conf/cgo/LattnerA04}, which helps to determine the set of affected functions.

 \noindent\textbf{Data-Flow Analysis}:
Variables propagate through operation statements (e.g., assignment or arithmetic), function parameters or return values.
In addition to explicit data-flow propagation, knob configuration  can also affect variables implicitly through conditional statements. 
Given the initial variables $v^k$, \sys performs data-flow analysis to identify the set of related variables (denoted as $\widetilde{V}^k$) propagated by $v^k$.

\noindent \textbf{Control-Flow Analysis}: 
Based on $\widetilde{V}^k$, \sys further identifies functions that are control-dependent on these variables.
The resulting set of controlled functions, denoted as $F^k$, serves as the basis for subsequent hypothesis formulation.

\begin{example}
Given the target knob \texttt{innodb\_log\_buffer\_size}, \sys first locates the initial program variable defined by the \texttt{Sys\_var\_*} data structures in the file \texttt{sys\_vars.cc}.
It then performs data-flow analysis to identify related variables, such as \texttt{log->buf\_size}, as shown in Figure~\ref{fig:control1}.
Finally, \sys identifies the functions that are control-dependent on these variables, such as \texttt{log\_buffer\allowbreak\_flush\_to\_disk()}, which are triggered when specific conditions involving the configuration knob are met.
\end{example}

After analyzing all knobs, \sys builds function–knob association mappings that record, for each function $f$, the set of controlling knobs $K^f$ such that adjusting any $k \in K^f$ influences the execution behavior of~$f$.
For instance, the execution of \texttt{log\_buffer\_flush\_to\_disk()} is governed by \texttt{innodb\_log\_buffer\allowbreak\_size}, while \texttt{buf\_read\_ahead\_random()} is controlled by both \texttt{innodb\allowbreak\_buffer\_pool\_instances} and \texttt{innodb\allowbreak\_random\allowbreak\_read\_ahead}.
This mapping allows \sys to identify knobs that control the performance-bottleneck functions, enabling more targeted tuning.

\subsection{Code-Driven Hypothesis Formulation}\label{sec:sem2}
Building upon the previously extracted data and control flows, \sys utilizes LLMs to analyze and summarize the influence of knob $k$ on the execution paths $F^k$ via iterative collaboration between a code retrieval agent and a code summary agent. 
Much like how developers read and reason about source code, understanding the effect of a knob often requires navigating across multiple files and symbols to resolve API calls and dependencies. To enable this reasoning, \sys parses the database codebase into an abstract syntax tree (AST) and provides two structured search APIs to facilitate efficient and targeted retrieval of relevant code fragments:
(1) \texttt{search\_function}(f) returns the implementation of the specified function;
(2) \texttt{search\_class}(cls) returns the signature of the specified class.

The retrieval process begins with the parent function that contains the control logic linking knob~$k$ to the controlled function(s) in $F^k$. 
From this context, the retrieval agent is guided to resolve any missing function or class references whose definitions are essential for deepening the understanding of how the knob influences system behavior.
To support this process, \sys adopts a stratified context search strategy~\cite{DBLP:conf/issta/0002RFR24}, which iteratively prompts the retrieval agent to determine which API calls are needed based on the current context. After the API invocations in a given stratum are executed, the newly retrieved code snippets are incorporated into the context. The retrieval agent is then instructed to assess whether the context is sufficient to analyze the influence of the configuration knob. Based on this assessment, the retrieval agent decides whether to (a) continue the iterative search process or (b) terminate retrieval when the code context is deemed complete.


\begin{example}
To understand how \texttt{innodb\allowbreak\_spin\allowbreak\_wait\allowbreak\_delay} affects $F^k$, the retrieval agent first receives the  code snippet of the parent function \texttt{rw\allowbreak\_lock\allowbreak\_x\allowbreak\_lock\allowbreak\_wait\allowbreak\_func()} and is prompted to assess whether the context is sufficient. 
In the first stratum, it invokes the \texttt{search\allowbreak\_function} API to retrieve the implementations of  unresolved functions within \texttt{rw\allowbreak\_lock\allowbreak\_x\allowbreak\_lock\allowbreak\_wait\allowbreak\_func()}, which are crucial for understanding the knob’s influence, such as \texttt{sync\allowbreak\_array\allowbreak\_wait\allowbreak\_event()}. In the second stratum, the context is augmented with the retrieved results. At this stage, the retrieval agent assesses that the context is sufficient and terminates the search process.
\end{example}

Once the retrieval agent determines that the context is sufficient or the query budget has been exhausted, the collected context is passed to the summary agent, which generates a structured reasoning chain that explains how the knob influences its associated functions and impacts overall database performance.
Specifically, to understand the underlying causal link, the summary agent first analyzes the influence path, including:  
(1) how changes in the knob value control the invocation of the target function and 
(2) the corresponding performance implications. 
Then, it formulates a tuning hypothesis based on the inferred causal relationships.

\begin{example}
Given the retrieved context of \texttt{innodb\_spin\allowbreak\_wait\allowbreak\_delay},  the agent first generates an explanation of its influence path:
``increasing this parameter makes \texttt{ut\_delay()} inject longer delays per spin, reducing context switches but increasing CPU busy-wait; decreasing it shortens or skips \texttt{ut\_delay()}, saving CPU but potentially increasing wakeups.''
The agent then formulates potential tuning strategies: ``if runtime profiling shows \texttt{ut\_delay()} consuming high CPU time, lowering \texttt{innodb\_spin\_wait\_delay} is recommended to reduce spin overhead; if \texttt{sync\_array\_wait\_event} dominates, increasing the delay helps avoid frequent context switch''.
\end{example}

Finally, \sys constructs a tuning hypothesis set  indexed by each knob~$k$ and its associated functions~$F^k$, which  records the promising knob adjustments  given different function behaviors.


\subsection{Tuning Rule Mining}\label{sec:num}

\blue{\marginnote{\textbf{R1W1,
R2W2,
R2D2, R4W2}}This section presents how \sys generates  quantitative  tuning rules based on semantic tuning hypotheses. 
These rules are, in essence, inductive generalizations over a collection of successful tuning observations under the semantic guidance.
The transformation involves two steps:}

\blue{\textbf{Semantic Sampling.}
\sys collects historical observations guided by semantic tuning hypotheses.
It first identifies important knobs using diagnosis-based knob selection (Section~\ref{sec:cycle}), retrieves their associated hypotheses, and applies the recommended configurations.
Each trial produces an observation that records: (i) the runtime context, (ii) the applied knob adjustment, and (iii) the observed performance impact. 
While a single successful observation could provide tuning guidance, its coverage is extremely limited because its contextual conditions are highly specific and rarely recur in future executions.}

\blue{\textbf{Inductive Rule Derivation.}
To obtain tuning guidance that generalizes across diverse contexts, \sys generates tuning rules (defined in Section~\ref{sec:ruled}) by mining the common patterns shared by successful observations. For example, if decreasing knob~$x$ by $\sim$10 consistently improves performance when the sampling rate of function~A appears at $2.5\%$, $3\%$, and $4\%$, \sys infers that the adjustment is beneficial within a broader interval such as $[2\%, 4\%]$. 
Using a customized Association Rule Mining (ARM) procedure (see Section~\ref{sec:arm}), \sys consolidates the consistent contexts and adjustment magnitude into tuning rules that achieve both high coverage and  high confidence across heterogeneous workloads.}



\subsubsection{Tuning Rule Definition}\label{sec:ruled}
\begin{definition} \textit{(Tuning Rules).}
A  tuning rule \(\mathcal{R}\) specifies how to adjust the knobs under certain system conditions.  
It can be expressed in the form $X\Rightarrow Y$, 
 where \(X\) denotes an itemset of context predicates that should be satisfied, and \(Y\) denotes an itemset of recommended knob adjustments when \(X\) holds true. 
An itemset may consist of a single element or multiple conjunctive items.
A tuning rule can be formally defined as:

\begin{equation}
\left(
\bigwedge_{k=1}^{K} p_k(w)
\ \land \
\bigwedge_{j=1}^{N} r_j \in [l_j, u_j]
\right)
\ \Longrightarrow \
\left(
\bigwedge_{i=1}^{M}
\theta_i : [\text{adjustment}]
\right)
\end{equation}
\[
\text{s.t.}\quad K + N\ge0 \quad \text{and} \quad M > 0.
\]
In this formulation, each \(p_k(w)\) denotes a workload predicate, such as whether workload type = OLTP, the number of concurrent threads > 10, or query arrival rate exceeding a threshold. Each \(r_j\) denotes the proportion of time spent in function \(j\), constrained within a range \([l_j, u_j]\). The symbols \(\theta_i\) denote the tunable knobs, and \([\text{adjustment}]\) describes the recommended modification to each knob, such as  decreasing \texttt{query\_timeout} by 10 seconds. 
This formulation indicates that when the context metrics meet the specified conditions, applying the recommended adjustments is expected to improve system performance.
\end{definition}

We adopt two metrics to evaluate the quality of tuning rules. To ensure that a rule is effective for future prediction,  it must generalize across a sufficiently large number of situations  (i.e., coverage) and demonstrate a high probability of making promising recommendations (i.e., confidence).

\begin{definition}
\textit{(Coverage).}
Let $\mathcal{H} = \{h_1, \dots, h_N\}$ denote a set of historical observations, and let $\mathcal{R}$ be a tuning rule expressed as $X \Rightarrow Y$. The \textbf{coverage} of $\mathcal{R}$ is defined as the proportion of observations in which the antecedent $X$ holds:
\[
\text{Coverage}(\mathcal{R}) = 
\frac{
  \left|\left\{
    h \in \mathcal{H}
    \mid
    X \text{ holds in } h
  \right\}\right|
}{
  \left|\mathcal{H}\right|
}.
\]
\end{definition}

\begin{definition}

\textit{(Confidence).}
Let $\mathcal{H}$ and $\mathcal{R}$ be as above. The \textbf{confidence} of $\mathcal{R}$ is defined as the proportion of observations in which  the consequent adjustment $Y$ improves the objective function $f$ when antecedent $X$ holds conditioned on all observations where the antecedent \(X\) holds and \(Y\) is applied:

\[
\text{Confidence}(\mathcal{R}) =
\frac{
  \left|\left\{
    h \in \mathcal{H}
    \mid
    \bigl(
      X \text{ and } Y\bigr)
    \text{ and }
    \bigl(
      f \text{ is improved }
    \bigr)  \text{ hold in } h
  \right\}\right|
}{
  \left|\left\{
    h \in \mathcal{H}
    \mid
    X \text{ and } Y \text{ hold in } h
  \right\}\right|
}.
\]

\end{definition}

\subsubsection{Tuning Rule Induction}\label{sec:arm}

To induce tuning rules with high coverage and confidence, we formulate it as  an association rule mining (ARM) problem~\cite{agrawal1993mining}, which aims to discover frequent co-occurrence patterns of contextual conditions and tuning actions that consistently lead to performance improvements.
 However, applying classical ARM techniques to database tuning is non-trivial due to two key challenges:
 \noindent\textit{C2.1. Data sparsity and generalization:} Benchmarking databases is costly, resulting in few observations. Rules learned from limited data risk overfitting and are hard to generalize across hardware and workload scales. For instance, increasing \texttt{buffer\_pool\_size} by 10\,GB may work on large-memory instances but fail on smaller ones. 
 \noindent \textit{C2.2. High-dimensional numerical data.} Knob values and runtime metrics are continuous, while traditional ARM handles categorical data. Naive discretization leads to a combinatorial explosion, high time complexity, and excessive memory usage~\cite{DBLP:journals/corr/abs-2307-00662}. 
To address these challenges, we propose a customized ARM pipeline for database tuning rule mining.

\noindent\textbf{Step1: Pairwise Data Augmentation.}
To overcome the sparsity of raw observations and enhance rule robustness, we construct pairwise records (i.e.,  transactions in ARM) from historical observations.
Given \(n\) historical observations, we enumerate all distinct pairs \((h_i, h_j)\). If \(h_i\) outperforms \(h_j\), we create an augmented record capturing: 
(1) the context predicates of \(h_j\), 
(2) the knob adjustments in \(h_i\) that transition from the lower- to higher-performing configuration. 
To localize promising actions, only pairs with performance improvement exceeding a threshold are retained.

\noindent\textbf{Step 2: Generalization-Oriented Encoding.}
We encode knob adjustments in a normalized and scale-invariant manner.
Memory-related knobs  are expressed as percentages of total memory to enhance comparability across  different  resource capacities. 
Knobs with wide numeric ranges are log-scaled, and those with smaller ranges are normalized using min-max scaling. Moreover, each adjustment is represented in both relative form (indicating the direction and magnitude of change) and absolute form (specifying the target value).  
These  encodings prevent the loss of meaningful tuning actions that may manifest differently across various scales and contexts,  improving the discovery of robust patterns.


\noindent\textbf{Step 3: Target-Constrained FP-Growth.}
To address the challenges of high-dimensional numerical data, we introduce performance-aware discretization and tuning-specific  constraints.
Specifically, we employ the Least Squared Ordinate-Directed Impact Measure (LSQM)~\cite{DBLP:conf/bigda/KaushikSPD21} to partition continuous features (e.g., knob adjustments and function sampling rates) into intervals that best capture their impact on performance.
\blue{\marginnote{\textbf{R4O1}}The discretization clusters continuous knob-adjustment magnitudes into performance-consistent intervals rather than preserving exact values.
For example, adjustments of 1\%, 12\%, and 30\% may fall into the same LSQM interval if they exhibit similar performance effects.
Therefore, the mined rules represent ranges of effective adjustments rather than relying on repeated occurrences of the same numeric magnitude.}
For pattern mining, we adopt FP-Growth~\cite{DBLP:journals/datamine/HanPYM04}, which features a compact FP-Tree structure and eliminates the need for candidate generation, compared to Apriori-based methods~\cite{DBLP:journals/corr/Heaton17a}.
We further enhance this process with \textit{target itemset pruning}, which restricts the search to candidate branches containing at least one tuning-related knob adjustment. 
Algorithm~\ref{alg:fp} outlines the workflow of the enhanced FP-Growth procedure.
It constructs a compressed prefix tree (FP-Tree) from the transaction database and recursively builds conditional FP-Trees to discover frequent itemsets. 
During the recursive exploration, branches that do not include any target items are pruned early (Lines~\ref{line:p1}--\ref{line:p2}), which significantly reduces the cost of constructing and traversing conditional FP-Trees.


\begin{algorithm}[t]
\caption{FP-Growth with Target Item Set Pruning}\label{alg:fp}
\KwIn{
  $Tree$: FP-Tree; 
  $prefix$: current prefix ;
  $target$: set of target items; 
  $min\_coverage$
}
\KwOut{
  Frequent itemsets containing any item in $target$
}

\If{$Tree$ is a single path}{
  \ForEach{combination in all combinations of $Tree$}{
    $candidate \gets prefix \cup combination$\;
    \If{$target \cap candidate \neq \emptyset$}{ output $candidate$\; }
  }
}
\Else{
  \ForEach{item $i$ in frequent items (ascending support)}{
    $nprefix \gets prefix \cup \{i\}$\;
    $P \gets$ extract\_conditional\_pattern\_base($Tree$, $i$)\;
    $P' \gets \emptyset$\;\label{line:p1}
    \ForEach{path $p$ in $P$}{
      \If{$(target \cap nprefix \neq \emptyset)$ \textbf{or} $(target \cap p \neq \emptyset)$}{
        $P' \gets P' \cup \{p\}$\;\label{line:p2}
      }
    }
    \If{$P'$ not empty}{
      $T' \gets$ build\_FP\_Tree($P'$, $min\_coverage$)\;
      FP-Growth($T'$, $nprefix$, $target$, $min\_coverage$)\;
    }
  }
}
\end{algorithm}

%% file: 5tuning.tex
\section{Configuration Recommender
}\label{sec:frame}


\subsection{Diagnosis-Based Knob Selection}\label{sec:detect}

Unlike existing knob selection approaches that primarily rely on extensive configuration observations~\cite{DBLP:conf/sigmod/AkenPGZ17, DBLP:conf/sigmod/CaiLZZZLLCYX22, DBLP:conf/kdd/FekryCPRH20} or workload-specific heuristics~\cite{DBLP:journals/pvldb/LaoWLWZCCTW24} (discussed in Section~\ref{sec:pre1}), \sys adopts a \emph{diagnosis-driven strategy} that enhances both the precision and interpretability of knob selection.
The key idea is to first diagnose bottleneck functions that contribute to performance degradation and then determine which knobs to tune based on explicit function--knob mappings (Section~\ref{sec:sem1}). 
Adjustments to these mapped knobs can propagate through program logic, affecting the behavior of the identified bottleneck functions, which enables targeted interventions to restore performance.

To achieve fine-grained identification of function-level performance bottlenecks, \sys leverages Linux performance analysis tools (e.g., \texttt{perf}~\cite{perf}) to  sample stack traces. 
While more sophisticated and lightweight profiling techniques such as eBPF~\cite{DBLP:conf/aipr2/DuS23} have been proposed, they are orthogonal to our focus and not explored in this work. 
Instead of relying on manual inspection of visualizations such as flame graphs~\cite{FlameGraph}, \sys automates bottleneck diagnosis by combining two complementary methods:

\textbf{Differential Profiling.} When representative normal and degraded performance states are available, \sys compares the sampling rates of each function across these two conditions. Functions that exhibit substantial variants in sampling proportion under degraded performance are flagged as likely contributors to performance issues (i.e., bottleneck functions).

\textbf{SHAP-Based Profiling.}
In scenarios where a clear normal baseline is difficult to define, \sys leverages SHAP (SHapley Additive exPlanations)~\cite{DBLP:conf/nips/LundbergL17} to quantify the marginal contribution of each function to performance degradation based on historical execution data. 
SHAP is particularly well-suited for bottleneck identification as it provides a principled, game-theoretic allocation of the performance discrepancy across all functions while rigorously accounting for inter-feature dependencies. This global attribution framework enables the precise isolation of functions that consistently contribute to performance degradation, even under strong feature correlations. 
Specifically, \sys fits a regression model on the  observed historical function sampling rates and the corresponding database performance. 
Given the current sampling rates, it computes SHAP values for each function and ranks them based on their negative contributions to overall database performance, which enables identification  of bottleneck functions without a predefined performance baseline.


 For each function~$f$ flagged as a potential bottleneck, \sys queries the function-knob mappings to identify its controlling knobs~$K^f$ as selected knobs. 
In addition, \sys matches the current runtime context against the antecedents of historical tuning rules; when a match is found, the knobs specified in the corresponding consequents are also incorporated into the set of selected knobs.


\subsection{Rule-Augmented Tuning}\label{sec:tune}

Directly relying on an LLM to generate configuration recommendations often results in incorrect or suboptimal outcomes due to hallucinations. To mitigate this issue, \sys incorporates verified rules to enhance recommendation reliability.

\subsubsection{Rule  Retrieval and Maintenance}
\sys retrieves relevant rules to guide the tuning process and continuously updates their statistics based on new observations to ensure reliability.

\noindent\textbf{Rule Retrieval.}
To identify candidate rules, \sys first gathers all rules whose antecedent predicates match the current runtime context,  referred to as \emph{relevant rules}.
Given the limited input window of the LLMs, \sys ranks these rules based on their \emph{expected improvement} and retrieves only the top-$k$ highest-scoring rules.

\begin{definition}
\textit{(Expected Improvement).}
Let $\mathcal{S}(\mathcal{R})$ denote the set of observations where applying $\mathcal{R}$ led to performance improvement, i.e.,
$
\mathcal{S}(\mathcal{R}) =
\left\{
h \in \mathcal{H}
\mid
(X \text{ and } Y) \text{ and } (f \text{ is improved}) \text{ hold in } h
\right\}$.
The \textbf{expected improvement} of rule $\mathcal{R}$ is defined as  the product of the average performance gain it achieves and the probability (confidence) that it leads to performance improvement:
\[
\text{EI}(\mathcal{R}) =
\frac{
\sum_{\substack{h \in \mathcal{S}(\mathcal{R})}}
\text{improvement}(h)
}{
|\mathcal{S}(\mathcal{R})|
}
\;\times\;
\text{confidence}(\mathcal{R}).
\]
\end{definition}

\noindent\textbf{Rule Maintenance.}
After the suggested configuration is applied and its performance impact is observed, \sys  updates the statistics of relevant rules.
Among these relevant rules, \sys identifies those whose prescribed knob adjustments match the actual applied adjustments, referred to as \emph{hit relevant rules}.
If the applied adjustment results in a performance improvement, both the numerator and denominator of the confidence metric for each hit relevant rule are incremented by 1; otherwise, only the denominator is incremented.
The \emph{expected improvement} of each hit relevant rule is then recalculated to incorporate the new observation.

\subsubsection{Rule-Augmented Prompt Generation}\label{sec:prompt}
While the tuning rules offer reliable guidance, they may not match the current conditions (i.e., no applicable rules are retrieved), or potentially better adjustments may remain undiscovered by the rules.
\sys therefore adopts a strategy that considers both \emph{exploitation} and \emph{exploration}:
(i) exploiting verified rules where applicable to ensure reliability, and
(ii) exploring recommendations guided by semantic hypotheses, which provide direction and help avoid overly aggressive or random exploration.
The  prompt mainly consists of three components: (1) Task instruction: A high-level description of the objective and expected behavior of the tuning agent;
(2) Task Information: Contextual information required to ground the recommendations, including the hardware, workload characteristics, identified bottleneck functions, and the selected knobs;
(3) Tuning Hypotheses and Rules: 
A set of semantic tuning hypotheses indexed by the selected knob and bottleneck functions, along with the retrieved rules.



\subsubsection{Hypotheses-to-Rule Life-cycle}\label{sec:cycle}
During the tuning process, \sys continuously transforms hypotheses into reliable rules.
Given the hypotheses associated with the selected knobs and the retrieved tuning rules, \sys suggests promising configurations for evaluation.
As \sys handles more tuning tasks, previously unseen knob-behavior patterns may emerge (e.g., new bottleneck-function interactions under varying workloads or hardware conditions).
To capture these patterns, \sys periodically applies the customized ARM pipeline to induce new tuning rules from accumulated observations.
Moreover, even when no applicable rules are available, \sys can still exploit hypotheses to discover superior configurations beyond the current rule set, which are subsequently validated and incorporated as new rules.



%% file: 6exp.tex
\section{Evaluations}
\subsection{Experiment Setting}\label{sec:exp-set}
\begin{figure*}[t]
    \centering
 {
    \includegraphics{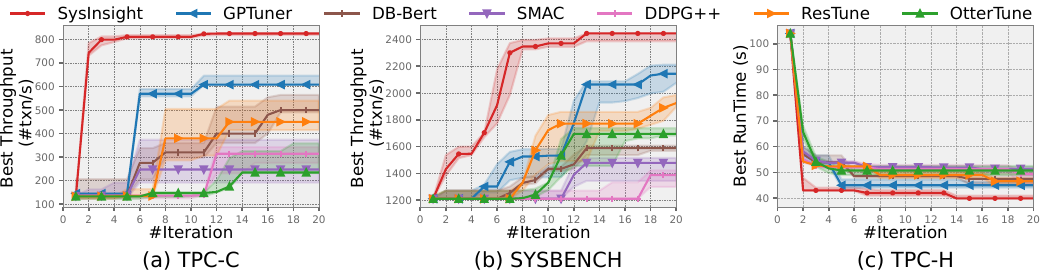}}
    \caption{Best Performance Over Iterations. For TPC-C and SYSBENCH, top-left is better. For TPC-H, bottom-left is better.}
    \label{fig:e2e}

\end{figure*}
\noindent\textbf{Workload.}
Following the setup of GPTuner~\cite{DBLP:journals/pvldb/LaoWLWZCCTW24}, we use TPC-H (OLAP type) with scaling factor 1 and TPC-C (OLTP type) with scaling factor 200.
TPC-C uses ten terminals with unlimited arrival rate and the implementations of benchmarks are from BenchBase~\cite{DifallahPCC13}.
Additionally, we employ SYSBENCH~\cite{SYSBENCH} in read-write mode, configured with 100 tables each containing 6 million rows.

\noindent\textbf{Hardware.}
We conduct  experiments on two cloud instance configurations: (i) Instance A, with 4 CPU cores and 16 GB RAM, used for TPC-H and SYSBENCH workloads; and (ii) Instance B, with 16 CPU cores and 64 GB RAM, used for TPC-C workloads.



\noindent\textbf{Baselines.}
We compare the performance of \sys with the state of the art tuning systems including manual-driven methods (GPTuner~\cite{DBLP:journals/pvldb/LaoWLWZCCTW24}, DB-Bert~\cite{DBLP:conf/sigmod/Trummer22}) and data-driven methods (SMAC~\cite{DBLP:conf/lion/HutterHL11}, DDPG++\cite{DBLP:journals/pvldb/AkenYBFZBP21}, ResTune~\cite{DBLP:conf/sigmod/ZhangWCJT0Z021}, OtterTune~\cite{DBLP:conf/sigmod/AkenPGZ17}).

\noindent\textbf{Offline Data Collection.}
To provide historical data for both transfer-based data-driven methods (i.e., ResTune\cite{DBLP:conf/sigmod/ZhangWCJT0Z021} and OtterTune~\cite{DBLP:conf/sigmod/AkenPGZ17}) and for \sys to mine tuning rules, 100 configurations are sampled per workload using Latin Hypercube Sampling~\cite{DBLP:conf/wsc/McKay92}.
To ensure a fair workload transfer scenario, all data collected under the target workload is strictly excluded from the tuning process.
For \sys, this means that the construction of tuning rules does not leverage any data from the target workload.

\noindent\textbf{Tuning Setting.}
The target system is MySQL 8.0.36 (Community Edition, GPL license). 
Following prior studies~\cite{DBLP:journals/pvldb/KanellisDKMCV22,DBLP:journals/pvldb/LaoWLWZCCTW24}, all baselines tune 44 knobs with MySQL default configuration as the starting point and optimize throughput for OLTP workloads and total workload execution time for OLAP workloads.
\blue{\marginnote{\textbf{R1W1,
R1W3,
R1D6}}Note that production deployments often adopt non-default configurations, such as  template settings (e.g., setting the buffer pool to 60\% of available memory). 
Thus, the absolute gains compared with production-tuned settings are expected to be smaller.}
We  run  three tuning sessions for each method with 20 iterations per session.
For \sys and GPTuner~\cite{DBLP:journals/pvldb/LaoWLWZCCTW24}, we employ GPT-4o-mini as the backend model.
As for DB-Bert, we adopt the authors' implementation which utilizes the pre-trained BERT model~\cite{DBLP:conf/naacl/DevlinCLT19}.
\blue{\marginnote{\blue{\textbf{R1D5, R4W1, R4W3}}}Due to space limitations, we include the additional evaluation on PostgreSQL in the technical report~\cite{SysInsightTR2025}, where the results are consistent with those observed on MySQL.}


\subsection{Efficiency Comparison}\label{sec:exp-eff}
Figure~\ref{fig:e2e} illustrates the performance of the best configurations found  by different approaches throughout the tuning iteration.
We report the median performance, with the interquartile range shown as shaded areas.
\sys consistently identifies the best-performing configurations with significantly fewer iterations across all benchmarks.
Compared with GPTuner, \sys converges to its best configuration on average $7.11\times$ faster while still delivering an average performance improvement of 19.9\%.

\subsubsection{Comparison with Manual-driven Methods.}
 GPTuner leverages  LLMs to extract structured knowledge of each knob  from documentation: (1) suggested range defined by \emph{min\_value} and \emph{max\_value}, (2) \emph{suggested\_values} and (3) \emph{special\_values}: 
Subsequently, it applies BO to explore a reduced and discretized search space defined by these ranges and values.
Compared with  data-driven methods, GPTuner effectively exploits prior knowledge from manuals to narrow the search space and accelerate convergence.
Nevertheless, the structured knowledge extracted  is limited and provides only moderate benefits.
 Among the 44 knobs, only 20 knobs have \emph{min\_value} or \emph{max\_value} differing from the MySQL defaults, 8 knobs include \emph{suggested\_values}, and 14 knobs have \emph{special\_values}. 
 
 The limitations mainly stem from two aspects.
First, the knowledge provided by manuals is coarse-grained and lacks the precision guidance for fine-grained optimization.
Second, for many knobs,  universal information applicable to all scenarios simply does not exist, making it impossible to extract meaningful ranges or values without considering the context.
According to the documentation of \texttt{innodb\_spin\_wait\_delay}, ``increasing this value can reduce CPU usage but may also increase response time, so it should be adjusted carefully based on system performance''.
While this qualitative insight is valuable, there is no well-defined range or suggested value that applies across all contexts, causing GPTuner to fail in extracting meaningful structured knowledge.

A similar limitation is observed in DB-Bert, which provides primarily context-agnostic guidance (with the exception of hardware-related factors) and derives meaningful knowledge for only 12 knobs.
In contrast, \sys extracts hypotheses for all 44 knobs directly from the source code and has discovered tuning rules for 33 knobs with high confidence.
This improvement arises because code-level analysis reveals fine-grained control information about each knob, providing richer and more context-aware references.
Moreover, the tuning rules defined by \sys explicitly consider how each knob should be adjusted under different runtime contexts,  enabling more targeted and effective optimization.

\subsubsection{Comparison with Data-driven Methods.}
SMAC is shown to be the best-performing BO-based method for DBMS tuning without transferring according to the evaluation study~\cite{DBLP:journals/pvldb/ZhangCLWTLC22}.
DDPG++ is an improved version of the Deep Deterministic Policy Gradient (DDPG) algorithm, originally introduced in CDBTune~\cite{DBLP:conf/sigmod/ZhangLZLXCXWCLR19} and enhanced in~\cite{DBLP:journals/pvldb/AkenYBFZBP21}.
However, within 20 iterations, SMAC and DDPG++ fail to identify high-quality configurations because they start the search process from scratch.
As for the baselines utilizing historical observations, OtterTune~\cite{DBLP:conf/sigmod/AkenPGZ17} identifies the most similar workload from a historical data repository through workload mapping and uses both matched data and target observations to train a surrogate model.
ResTune~\cite{DBLP:conf/sigmod/ZhangWCJT0Z021} fits multiple base learners on observations from each source task and combines their predictions into a meta-learner via dynamic weighting.
Compared with OtterTune, ResTune achieves better performance by mitigating negative transfer.
However, both methods fundamentally rely on the similarity of performance surfaces across configurations, and their effectiveness degrades when the target workload is insufficiently similar to the source workloads.
In contrast, \sys leverages insights from database internals---derived from source code and execution context---that capture generalizable relationships between system behaviors and knob adjustments (e.g., observing frequent page misses suggests increasing the buffer pool), enabling robust and interpretable optimization beyond specific workloads.

\begin{figure}[t]
    \centering
 {
    \includegraphics{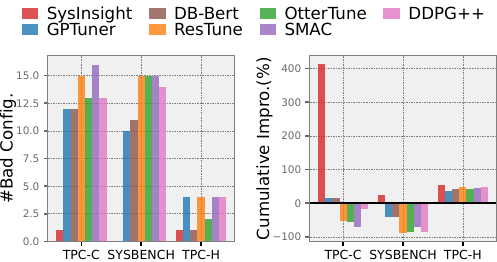}}
    \caption{Reliability Comparison. Left figure shows \#bad configurations (low is better). Right figure shows cumulative improvement over the default (higher is better).}
    \label{fig:unsafe}
\end{figure}

\subsection{Reliability Comparison}\label{sec:exp-reliable}
During tuning, it is crucial not only to eventually find the best-performing configuration but also to ensure that intermediate recommended configurations do not degrade system performance, referred  as  \textit{reliability}.
If the tuning algorithm explores too many poor configurations, the cumulative performance may fall below the default baseline, resulting in an overall performance loss.
To quantify reliability, we adopt two complementary indicators: (1)\#Bad Configurations: The number of recommended configurations that perform worse than the default. (2)Cumulative Improvement: The cumulative performance relative to the default during the tuning process. 
For OLTP workloads, it is measured by the total number of processed transactions (\#txn), while for OLAP workloads, it is evaluated based on the total accumulated query execution time.

Figure~\ref{fig:unsafe} presents the two metrics for each method across the three workloads. 
\sys consistently achieves the fewest \#Bad configurations, with only one instance reported in both TPC-C and SYSBENCH.
 For cumulative improvement, a positive value indicates that the tuning process yields a net performance gain over the default, while a negative value implies that the algorithm's exploration leads to overall performance degradation. 
SysInsight again demonstrates the best results, achieving a remarkable 412.8\% improvement on TPC-C and maintaining positive gains on both SYSBENCH (24.2\%) and TPC-H (19.8\%).
By contrast, other methods fail to amortize the cost of exploration on TPC-C and SYSBENCH. 
In TPC-H, however, substantial gains can be realized by enlarging buffer pool size, which partly explains the relatively better performance of the baseline methods.

\begin{table*}[t]
\centering
\caption{Examples of  Tuning Hypotheses and Rules.
 Each column presents a representative case where \sys selects tuning knobs based on bottleneck functions, retrieves  tuning hypothesis and rules of high confidence score.}\label{tbl:hy}
\begin{tabularx}{\textwidth}{|l|X|X|X|}

\hline
\textbf{Function} 
& \texttt{buf\_LRU\_get\_free\_block} 
& \texttt{srv\_purge\_coordinator\_thread} 
& \texttt{ut\_delay}, \texttt{sync\_array\_wait\_event} \\
\hline
\textbf{Associate Knob} 
& \texttt{innodb\_buffer\_pool\_size} 
& \texttt{innodb\_purge\_batch\_size} 
& \texttt{innodb\_spin\_wait\_delay} \\
\hline
\textbf{Causal Link} 
& \texttt{innodb\_buffer\_pool\_size} controls memory space for pages in \texttt{buf\allowbreak\_LRU\allowbreak\_get\_free\_block()}. A small pool triggers frequent evictions and dirty page flushes.
& \texttt{innodb\_purge\_batch\_size} controls the purge batch size in \texttt{srv\allowbreak\_purge\allowbreak\_coordinator\_thread()}. A small value slows undo log cleanup and lengthens MVCC chains; a large value can cause I/O spikes.
& \texttt{innodb\_spin\_wait\_delay} controls the spin duration in \texttt{ut\allowbreak\_delay()}. A short delay leads to frequent context switches (\texttt{sync\_array\_wait\allowbreak\_event()}); a long delay wastes CPU in spin-lock polling. \\
\hline
\textbf{Hypothesis} 
& If \texttt{buf\allowbreak\_LRU\allowbreak\_get\allowbreak\_free\allowbreak\_block()} call is frequent, increase \texttt{innodb\allowbreak\_buffer\allowbreak\_pool\allowbreak\_size}. 
& If \texttt{srv\_purge\_coordinator\_thread()} call is frequent and undo log buildup slows snapshot reads (\texttt{row\allowbreak\_search\allowbreak\_mvcc}()), increase the knob.
& If \texttt{sync\allowbreak\_array\allowbreak\_wait\allowbreak\_event()} call is frequent, increase the knob; if \texttt{ut\_delay()} call is frequent but lock contention remains, decrease it. \\
\hline
\textbf{Tuning Rule} 
& $r(\texttt{buf\_LRU\_get\_free\_block})$ $>$ 16\% $\Rightarrow$ increase \texttt{innodb\allowbreak\_buffer\allowbreak\_pool\allowbreak\_size} by (0.16Mem, 0.50Mem]
& $r(\texttt{srv\_purge\_coordinator\_thread})$ > 5\% $\land$ $r(\texttt{row\_search\_mvcc}) > 10\% \Rightarrow$ increase \texttt{innodb\allowbreak\_purge\allowbreak\_batch\_size} by (0,100]
& $r(\texttt{sync\_array\_wait\_event}) > 3\% \Rightarrow$ increase \texttt{innodb\allowbreak\_spin\allowbreak\_wait\allowbreak\_delay} by (0,10] \\
\hline
\textbf{Confidence} 
& 0.87 & 0.85 & 0.74 \\
\hline
\end{tabularx}
\end{table*}


\subsection{Case Study}\label{sec:exp-case}
To further demonstrate the effectiveness of \sys, we conduct a case study using the TPC-C benchmark. 
We \blue{analyze the human effort required to formulate tuning hypothesis,}  
illustrate how the identified bottlenecks guide knob selection, and how the retrieved tuning hypotheses and rules provide effective guidance.

\begin{figure}[t]
    \centering
    \begin{subfigure}[t]{0.445\linewidth}
        \centering
        \includegraphics[width=\linewidth]{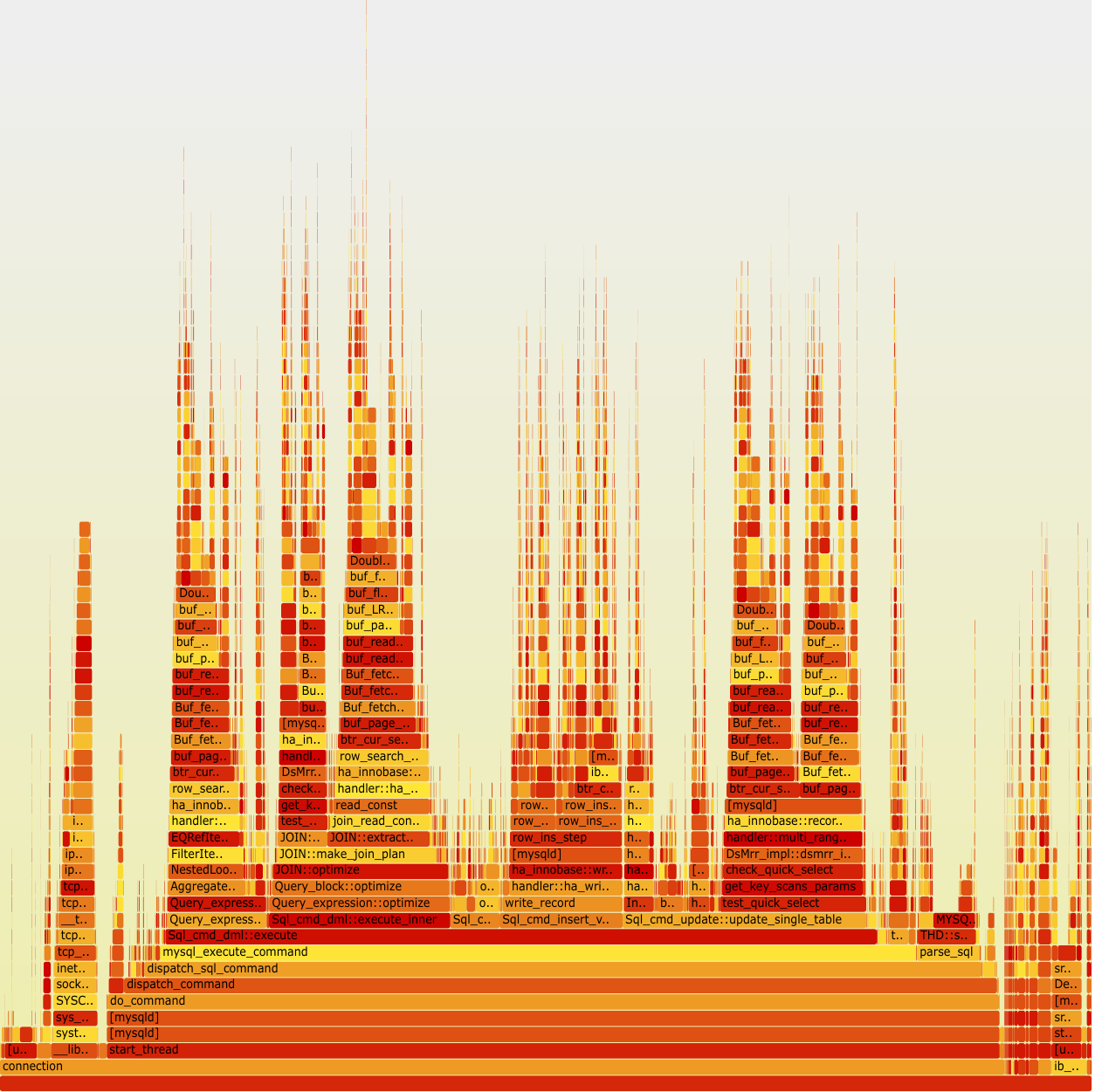}
        \caption{Default Configuration.}
        \label{fig:flame_default}
    \end{subfigure}
    \hfill
    \begin{subfigure}[t]{0.45\linewidth}
        \centering
        \includegraphics[width=\linewidth]{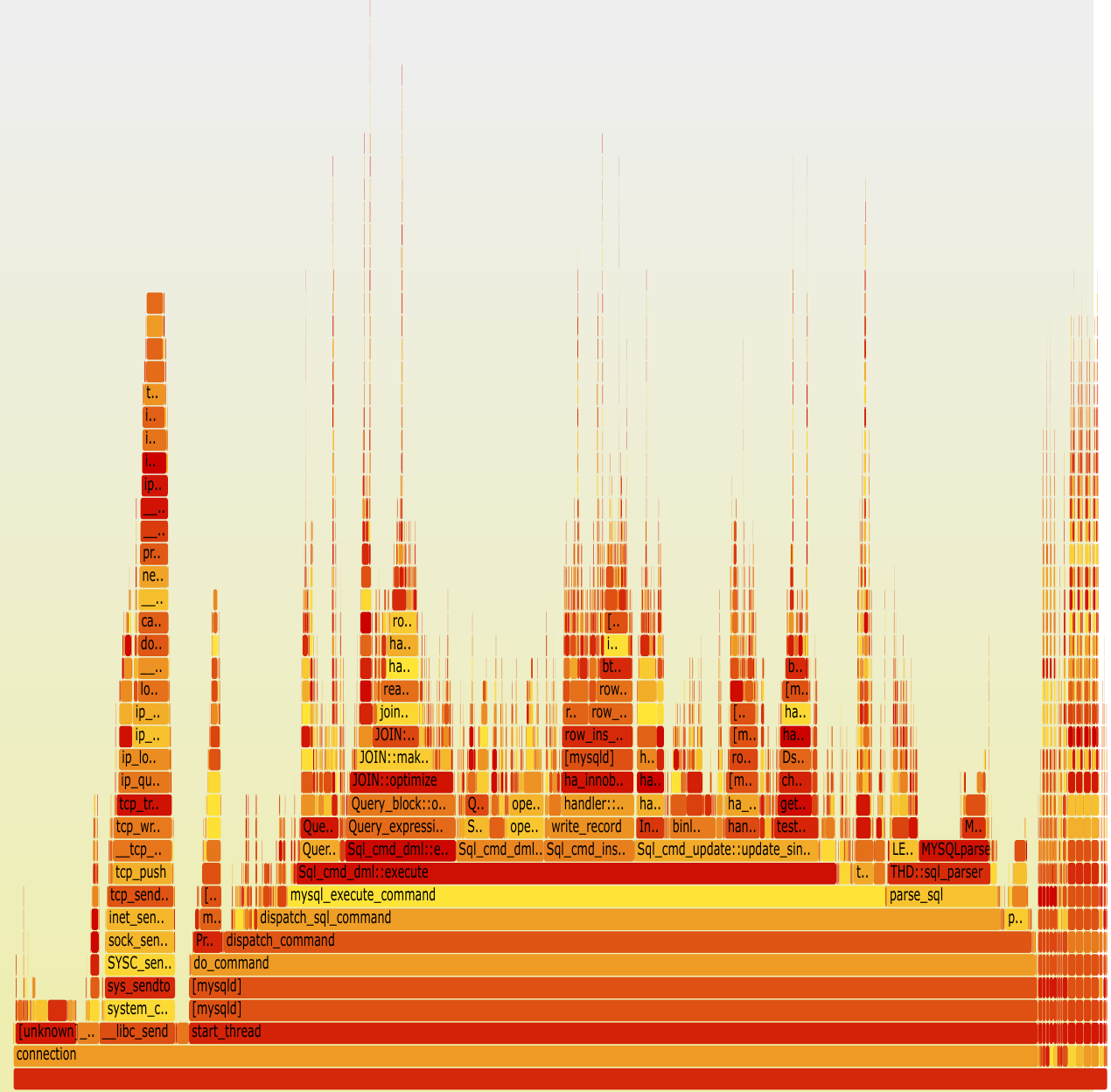}
        \caption{Best Configuration.}
        \label{fig:flame_best}
    \end{subfigure}
    \caption{Different System Behaviors Shown in Flame Graphs. The width of each function represents its runtime cost.}
    \label{fig:flame_compare}
\end{figure}

\textbf{\blue{\marginnote{R2W1, R2D1}Tuning Hypothesis  Formulation.}}
\blue{
We analyze the human effort required for deriving the  tuning hypothesis  for \texttt{innodb\_buffer\allowbreak\_pool\_size}.
Static analysis identifies 42 functions that are control-dependent on this knob.
A database expert spent approximately one full day tracing the knob’s propagation and understanding its impact paths.
For each of these 42 functions, it was typically necessary to examine an additional 10 surrounding or dependent functions to fully understand their behavior.
This further illustrates that even analyzing a single knob involves considerable effort, and extending such analysis to dozens of knobs would be highly labor-intensive.}

\textbf{Diagnosis-based Knob Selection.} 
As shown in Figure~\ref{fig:flame_compare},
under the default configuration, wide I/O hotspots dominate the runtime. 
\sys identifies functions such as \texttt{buf\_page\_get\_gen} and \texttt{buf\_LRU\_get\_free\_block} as  bottlenecks, indicating that the I/O path (disk page loading and LRU page allocation) is the key performance constraint. 
Additionally, other critical bottleneck functions such as \texttt{srv\_purge\_coordinator\_thread} (responsible for undo log cleanup) and \texttt{sync\_array\_wait\_event} (related to lock contention and spin-wait synchronization) are also identified. 
By querying the function–knob mapping, \sys selects \texttt{innodb\_buffer\allowbreak\_pool\_size}, \texttt{innodb\_purge\_batch\_size} and \texttt{innodb\_spin\_wait\allowbreak\_delay}  that these bottleneck functions are control-dependent on.


\textbf{Rule-Guided Configuration Tuning.}
Based on the bottleneck diagnosis, \sys retrieves the corresponding hypotheses and tuning rules for the selected knobs from the knowledge library, as shown in Table~\ref{tbl:hy}. 
Those rules  provide interpretable and context-aware guidance by linking observed  bottlenecks to promising configuration adjustments, each accompanied by a confidence score that indicates the reliability. 
For example, while manual guidance suggests that setting \texttt{innodb\_buffer\_pool\_size} to 70--80\% of system memory yields good performance (as also adopted by GPTuner), \sys leverages validated high-confidence rules to dynamically adjust this knob based on runtime function-level signals (e.g., \texttt{buf\_LRU\_get\_free\_block}). 
As a result, \sys recommends a buffer pool size of around 50\% of available memory, which achieves superior performance by reducing I/O overhead while preserving sufficient memory for other critical components.


\begin{figure*}[htbp]
    \centering
    \begin{minipage}{0.69\textwidth}
        \centering
        \includegraphics[width=\textwidth]{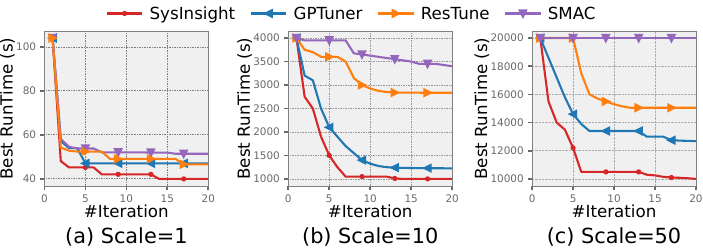}
        \caption{Effect of Database Size on Tuning Performance (bottom-left is better).}
        \label{fig:data}
    \end{minipage}
    \hfill
    \begin{minipage}{0.27\textwidth}
        \centering
        \includegraphics[width=\textwidth]{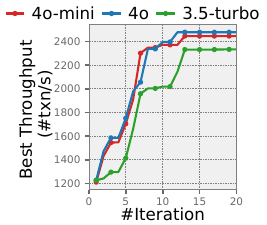}
        \caption{Effect of  Language Models  (top-left is better).}
        \label{fig:llm}
    \end{minipage}
    \vspace{-0.3em}
\end{figure*}

\subsection{Scalability Study}
We study the impact of database size on tuning performance by conducting experiments on TPC-H workloads with scale factors of 1, 10, and 50. 
We compare \sys with three representative baselines from different categories: 
(i) GPTuner (manual-driven), 
(ii) ResTune (data-driven with transfer learning), 
and (iii) SMAC (data-driven without transfer learning). 
Figure~\ref{fig:data} shows the best runtime achieved versus the number of tuning iterations under different scales. 
As the database size grows, the absolute runtime of all methods increases, but \sys consistently achieves the fastest convergence and the lowest final runtime. 
The performance of data-driven methods degrades noticeably as the database grows, and SMAC fails to find any configuration better than the default  at a scale factor of 50.
This degradation occurs because larger datasets result in a sparser promising performance landscape, making it harder for data-driven methods to identify good configurations.

Unlike these  baselines which rely solely on a single statistical metric (e.g., total execution time for TPC-H) to explore the configuration space, \sys leverages code-level diagnostics and causal hypotheses to conduct targeted tuning, thereby mitigating the challenges of sparse performance landscapes as database size grows.
For example, \sys observes a bottleneck that \texttt{TableScanIterator::Read} incurs a high cost, while \texttt{buf\_read\allowbreak\_ahead\allowbreak\_linear} is rarely invoked, indicating that the linear read-ahead mechanism is underutilized. 
From the function-knob mapping,  \texttt{innodb\_read\_ahead\_threshold} is identified as the controlling knob, and \sys lowers this threshold to activate read-ahead logic more aggressively, improving sequential scan performance.
Similarly, \sys detects the high cost of hash join operations (e.g., \texttt{HashJoinIterator::BuildHashTable}) and doubles its associated knob \texttt{join\_buffer\_size} from 256KB to 512KB, which allows more intermediate join results to be retained in memory and accelerating hash table construction.

\subsection{Analysis of \sys}

\subsubsection{Overhead Analysis}
\sys incurs offline overhead from tuning hypothesis formulation and rule mining. 
Hypotheses are generated once per DBMS version, requiring 45 hours for function–knob extraction and 4 hours (\$5) for LLM-based generation using 1.5M input and 121K output tokens.  
All results are cached for reuse across future tuning tasks.
The online tuning overhead consists of:  
(1) the time taken by the optimizer to suggest the next configuration, and  (2) the workload evaluation time.  
On average, \sys spends about 26 seconds to produce the next configuration, including 0.03 seconds for diagnosis-based knob selection, 0.3 seconds for hypothesis and rule retrieval, and 25 seconds for the tuning agent’s recommendation.   

\begin{figure}[t]
    \centering
 {
    \includegraphics{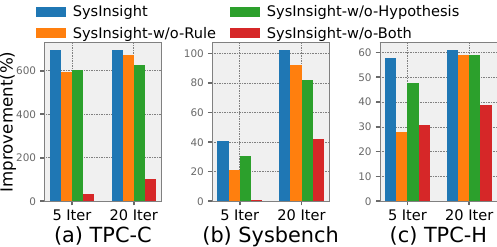}}
    \caption{Ablation Study on Tuning  Hypotheses and Rules.}
    \label{fig:ablation}
\end{figure}

\subsubsection{Effect of Different Language Models.}
As shown in Figure~\ref{fig:llm}, we compared the performance of different LLMs, including {gpt-4o}, {gpt-4o-min}, and {gpt-3.5-turbo}, which are used as the backend model for tuning agents. As the capability of the underlying model improves, \sys achieves better results, indicating that \sys can effectively leverage the power of LLMs. Among them, {gpt-4o-min} performs comparably to {gpt-4o} while incurring only about 6\% of the cost per token.
Therefore, we adopt gpt-4o-min as the default backend model in \sys to achieve a favorable balance between performance and cost.

\subsubsection{Effect of Hypotheses and Rules}
Figure~\ref{fig:ablation} shows the  performance improvement against the default configuration of \sys when removing hypothesis retrieval, rule retrieval, or both, measured after 5 and 20 iterations. We observe that removing both components leads to a significant performance drop, while removing only one has a relatively smaller impact, indicating that the two components are complementary. Moreover, \sys-w/o-hypothesis performs better in early stages (5 iterations), suggesting that rules can quickly provide good configurations. In contrast, \texttt{\sys-w/o-rule} achieves better performance in  later stages (20 iterations), indicating that hypotheses help explore potentially better configurations.

\begin{figure}[t]
    \begin{minipage}{0.235\textwidth}
        \centering
        \includegraphics{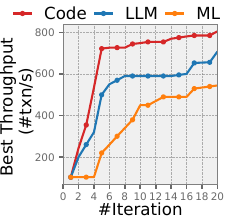}
        \caption{Knob Selection: We compare code-based, LLM-based and ML-based Methods.}
        \label{fig:knob}
    \end{minipage}
    \hfill
    \begin{minipage}{0.23\textwidth}
        \centering
        \includegraphics{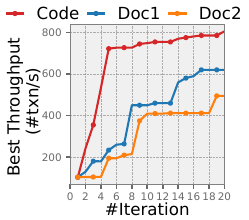}

        \caption{Knowledge Source: We compare code-driven and manual-driven methods.}
        \label{fig:source}
    \end{minipage}
\end{figure}

\subsubsection{Comparison of Knob Selection Methods.}\label{sec:exp-know}
We compare different knob selection methods by replacing the knob selection component of \sys. To isolate the effect of knob selection, we remove the rule retrieval module of \sys, which otherwise implicitly influences knob   choices. The compared methods are:
(1) Code-based: Our methods that identifies bottleneck functions and leverages static code analysis to find the associated knobs for tuning;
(2) LLM-based: the approach used by GPTuner~\cite{DBLP:journals/pvldb/LaoWLWZCCTW24}, prompts LLMs  to select important knobs from the candidate set based on workload information;
(3) ML-based: the approach used by Tuneful~\cite{DBLP:conf/kdd/FekryCPRH20}, which begins with tuning all knobs and removes 40\% of the less important knobs every 5 iterations  using Gini importance ranking derived from prior observations.
Figure~\ref{fig:knob} shows the results on the TPC-C workload. The code-based method achieves the best performance due to its accurate identification bottleneck functions and associate knobs through code analysis. 
In contrast, the ML-based method performs worst, as the limited number of observations is insufficient to yield reliable knob importance rankings.

\subsubsection{Comparison of Different Knowledge Sources.}
We investigate the impact of different knowledge sources on \sys by replacing its code-derived tuning hypotheses with alternative sources:  structural knowledge from GPTuner~\cite{DBLP:journals/pvldb/LaoWLWZCCTW24} (denoted by Doc1) and  
 tuning hints from DB-Bert~\cite{DBLP:conf/sigmod/Trummer22} (denoted by Doc2).  
To avoid interference, the rule retrieval module of \sys is disabled during this comparison.  
As shown in Figure~\ref{fig:source} on  TPC-C, our code-based source  achieves the best performance, providing context-aware and fine-grained knowledge compared with the structural knowledge or tuning hints extracted from database manuals.

%% file: 7conclusion.tex
\section{Conclusion and Outlook}

We presented \sys, a code-driven tuning system that derives fine-grained, context-aware, and verifiable tuning insights directly from DBMS source code.
Unlike prior works, \sys takes a concrete step toward white-box tuning by explicitly leveraging system internals to guide the optimization process.
Extensive experiments  demonstrate its efficiency, reliability and scalability.
While \sys focuses on code-driven tuning, we acknowledge that other sources---such as text documentation and historical observations---can be complementary. 
As future work, we plan to explore hybrid strategies that unify these sources to further enhance tuning agents' adaptability and effectiveness.


%% file: 8ack.tex
\section{ACKNOWLEDGMENTS}
This work was supported by the National Natural Science Foundation of China under Grant Nos. 62502522, 62441230, 62472429, 62461146205, the new faculty start-up fund of RUC (Grant Nos. 202530153 and 202430021) and sponsored by CCF-ApsaraDB Research Fund. 
We thank the anonymous reviewers for their constructive suggestions, which help improve the quality of this paper.